\documentclass{webofc}
\usepackage{listings}
\usepackage[draft]{minted}
\usepackage[caption=false]{subfig}
\usepackage{hyperref}
\usepackage{graphicx}
\usepackage[utf8]{inputenc}
\usepackage{tikz}

\usepackage[varg]{txfonts}

\newminted{cpp}{gobble=2, numbersep=3pt, escapeinside=||, xleftmargin=5mm, framesep=2mm, frame=leftline, baselinestretch=1, fontsize=\scriptsize, linenos=true}
\newmintinline{cpp}{}

\makeatletter
\AtBeginEnvironment{minted}{\dontdofcolorbox}
\def\dontdofcolorbox{\renewcommand\fcolorbox[4][]{##4}}
\makeatother

\usetikzlibrary{shadows, arrows, fit, positioning, matrix, shapes}
\usetikzlibrary{decorations.markings}

\definecolor{my_yellow}{RGB}{255, 253, 217}
\definecolor{my_orange}{RGB}{255, 127, 0}
\definecolor{my_lightblue}{RGB}{105, 186, 249}
\definecolor{my_purple}{RGB}{150, 154, 219}
\definecolor{my_green}{RGB}{90, 194, 160}

\tikzset {
  bigbox/.style = {draw, thick, fill=gray!10, rounded corners, rectangle},
  box/.style = {draw, thick, minimum height=0.8cm, minimum width=1.5cm, rounded corners, rectangle, fill=white, anchor=south},
  model/.style = {draw, thick, fill=white, text centered, minimum height=3em, minimum width=4em, rounded corners, drop shadow},
  user/.style = {draw, thick, ellipse, fill=white, text centered, minimum height=3em, minimum width=5em, drop shadow},
  line/.style = {->, thick, color=black, shorten <=2pt, shorten >=2pt, >=stealth'},
  plain/.style = {minimum width=1em},
  arcnode/.style 2 args={
    decoration={
                 raise=#1,             
                 markings,   
                 mark=at position 0.5 with {\node[inner sep=0] {#2};}
            },
            postaction={decorate}
    }
}

\pgfdeclarelayer{background}
\pgfdeclarelayer{foreground}
\pgfsetlayers{background,main,foreground}

\begin{document}

\title{Optimizing Frameworks’ Performance Using C++ Modules Aware ROOT}

\author{\firstname{Yuka}
\lastname{Takahashi}\inst{1,2}
\fnsep\thanks{\email{yuka.takahashi@cern.ch}}
        \firstname{Vassil}
        \lastname{Vassilev}\inst{1}\fnsep\thanks
        {\email{vvasilev@cern.ch}}
        \firstname{Oksana}
        \lastname{Shadura}\inst{3}
        \fnsep\thanks{\email{oksana.shadura@cern.ch}} \and
        \firstname{Raphael}
        \lastname{Isemann}\inst{2,4}
        \fnsep\thanks{\email{isemann@student.chalmers.se}}
}

\institute{Princeton University, Princeton, New Jersey 08544, United States
\and
           CERN, Meyrin 1211, Geneve, Switzerland
\and
           University of Nebraska Lincoln, 1400 R St, Lincoln, NE 68588, United States
\and
           Chalmers University of Technology, Chalmersplatsen 4, 41296 Göteborg, Sweden
}

\abstract{%
ROOT is a data analysis framework broadly used in and outside of High Energy Physics (HEP). Since HEP software frameworks always strive for performance improvements, ROOT was extended with experimental support of runtime C++ Modules. C++ Modules are designed to improve the performance of C++ code parsing. C++ Modules offers a promising way to improve ROOT's runtime performance by saving the C++ header parsing time which happens during ROOT runtime. This paper presents the results and challenges of integrating C++ Modules into ROOT.
}
\maketitle
\section{Introduction}
\label{intro}

The core part of HEP data analysis framework ROOT \cite{root} is the C++ interpreter \texttt{Cling}. Cling is build on the top of the C++ compiler Clang and it allows users to enable interactive ROOT sessions. It also serves as a backend for ROOT's language bindings such as ROOT Python extension module PyROOT.

In order to offer these features, Cling has to parse source code during runtime. This includes not only the code manually entered by the user from a command line, but also a multitude of the header files coming from libraries and frameworks. The header file parsing can negatively affect ROOT's performance as it consumes notable amounts of memory and CPU time.

Since the inefficient header parsing in C++ is a well-known issue, it was introduced C++ Modules, delivering a compact, efficient representation of header files. Modules are becoming a promising technology for C++ community and since the adoption of C++ Modules into ROOT was already proposed earlier in paper \cite{vassil-paper} and we implemented them in ROOT and its interpreter Cling. This allowed us to avoid the expensive parsing of headers and improve ROOT's runtime performance.

\section{Background}
\label{background}

Even before C++ Modules adoption, ROOT has been heavily optimized and exploits several mechanisms to improve the performance of the interpreter. In this section we will discuss existing mechanisms in the context of C++ Modules: ROOT precompiled header (PCH), ROOTMAP files (ROOTMAP) and RDICT files.

\subsection{Optimizing ROOT using a PCH}
\label{pch}
ROOT ships with a precompiled header (PCH) available for a subset of ROOT's components. The PCH the CPU and memory cost for heavily used ROOT libraries. The PCH technology has been well-understood for decades. It is an efficient on-disk representation of the state of the compiler after parsing a set of headers. It can be loaded before starting the next instance to avoid doing redundant work. However, this approach limits compiler to use only a single PCH, as it's usually too involved to merge multiple compiler states loaded from different PCH files. ROOT’s dictionary generator, \textit{rootcling}, generates one PCH file at build time, which is loaded on startup and ROOT proceeds to lazily evaluate the code from the file when needed.

The PCH is by design monolithic and not extensible. It introduces problems for third party libraries who want to improve the performance of their code with a PCH. They cannot provide a second PCH file with their code as this is not possible by design. For ROOT developers the PCH brings the problem that changing a single header requires the regeneration of the whole PCH file.

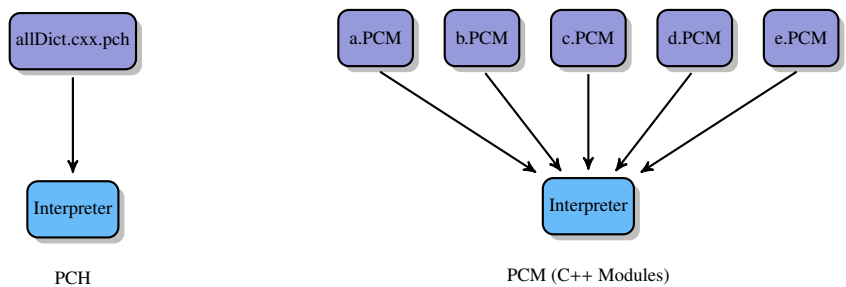
\begin{figure}[!h]
  \centering
  \begin{tikzpicture}[outer sep=0.05cm, node distance=0.8cm, scale=0.7, transform shape]
        
    \node[model, fill=my_purple, name=pch] (pch) {allDict.cxx.pch};
    \node[model, fill=my_lightblue, name=interpreter, below=2cm of pch] (interpreter) {Interpreter};

    \draw[line, ->] (pch.south) -- (interpreter);
    
    \node [below=1cm, align=flush center,text width=8cm] at (interpreter) { PCH };
  \end{tikzpicture}
  \hfill
  \begin{tikzpicture}[outer sep=0.05cm, node distance=0.8cm, scale=0.7, transform shape]
        
    \node[model, fill=my_purple, name=a] (a) {a.PCM};
    \node[model, fill=my_purple, name=b, right=0.5cm of a] (b) {b.PCM};
    \node[model, fill=my_purple, name=c, right=0.5cm of b] (c) {c.PCM};
    \node[model, fill=my_purple, name=d, right=0.5cm of c] (d) {d.PCM};
    \node[model, fill=my_purple, name=e, right=0.5cm of d] (e) {e.PCM};

    \node[model, fill=my_lightblue, name=interpreter, below=2cm of c] (interpreter) {Interpreter};

    \draw[line, ->] (a.south) -- (interpreter);
    \draw[line, ->] (b.south) -- (interpreter);
    \draw[line, ->] (c.south) -- (interpreter);
    \draw[line, ->] (d.south) -- (interpreter);
    \draw[line, ->] (e.south) -- (interpreter);
    
    \node [below=1cm, align=flush center,text width=8cm] at (interpreter) { PCM (C++ Modules) };

  \end{tikzpicture}
  \caption{Comparison of PCH and C++ Modules in ROOT. The content of a single PCH can be separated to multiple PCMs, which makes PCMs modular.}
  \label{fig:pchandPCM}
\end{figure}

A less restrictive alternative to PCH files is C++ Modules (PCM files). Unlike a PCH, PCM files are designed to be separable, so that multiple PCMs can be attached to the same interpreter or compiler simultaneously. This means that it is possible to split up the content of a single PCH into multiple PCMs as shown in Figure \ref{fig:pchandPCM}. Also, it is possible to rebuild only a subset of all attached PCMs. In ROOT, a single PCM file usually corresponds to a single library.

The implementation of C++ Modules in ROOT is based on Clang's implementation, which is called as an API from ROOT. Clang uses configuration files called Module Maps for defining the contents of a C++ Module. To stay consistent with Clang, ROOT also uses Module Map files to configure Module contents.

\subsection{Optimizing third-party code with ROOTMAP and RDICT}
\label{dictgen}

ROOTMAP files are used by ROOT to map unknown symbols and identifiers to libraries. When encountering an unknown symbol or identifier, ROOT uses ROOTMAP files to load the corresponding library. An advantage of this approach is that ROOTMAP files allow ROOT to only parse headers of libraries that are actually used. The expensive header parsing only happens when it is necessary.

RDICT files efficiently store information needed for serialization and deserialization of data types. They allow ROOT to perform IO operations on these data types without loading the respective code from the PCH or parsing the respective library headers.

\begin{listing}[h]
    \noindent
    \begin{minipage}[h]{\textwidth}
    \begin{cppcode*}{}
    // Foo.h
    namespace foo { struct bar{}; }
    struct S{};

    // libFoo.rootmap
    { decls }
    namespace foo { }
    struct S;
 
    [ libFoo.so ]
    # List of selected classes
    class bar
    struct S

    // G__Foo.cxx (aka libFoo dictionary)
    namespace {
      void TriggerDictionaryInitialization_libFoo_Impl() {
        static const char* headers[] = {"Foo.h"}
        // More scaffolding
        extern int __Cling_Autoloading_Map;
        namespace foo{struct __attribute__((annotate("$ClingAutoload$Foo.h"))) bar;}
        struct __attribute__((annotate("$ClingAutoload$Foo.h"))) S;
       // More initialization scaffolding.
    }
    \end{cppcode*}
    \end{minipage}
    \caption{The example of ROOT dictionary for {\it libFoo}. {\it Foo.h} is the header file which contains the definition. {\it libFoo.rootmap} is generated by rootcling, and it contains the forward declaration of {\it Foo.h}. {\it G\_\_Foo.cxx} is injected in the {\it libFoo} object file, and is called when {\it libFoo} is being loaded to ROOT.}
    \label{list:foo}
\end{listing}

At startup, ROOT will locate all files with extension {\it *.rootmap}. It parses the code in Listing \ref{list:foo} Line 6 \{decls\} section and creates an internal map for the entities defined in Listing \ref{list:foo} Line 10 [libFoo.so] section. Upon seeing an unknown identifier, the implementation searches in the database if this is a known entity.

\begin{listing}[h]
    \noindent
    \begin{minipage}[h]{\textwidth}
    \begin{cppcode*}{linenos=true}
    root [] S *s;           // Does not require a definition.
    root [] foo::bar *baz1; // Does not require a definition.
    root [] foo::bar baz2;  // Requires a definition.
    \end{cppcode*}
    \end{minipage}
    \caption{Illustrative example for usage of the ROOT dictionary contents.}
    \label{list:prompt}
\end{listing}

Listing \ref{list:prompt} shows the efforts which ROOT does to avoid parsing redundant code. {\it S} is defined in Line 3 in Listing \ref{list:foo} and {\it foo::bar} is defined in Line 2 in Listing \ref{list:foo}. Line 1 does not require a definition and the forward declaration consumed at the initialization time is sufficient, so the parsing of {\it Foo.h} is not required. The behavior of Line 1 in Listing \ref{list:prompt} is equivalent to Line 1 and 2 in Listing \ref{list:line2}.

\begin{listing}[h]
    \noindent
    \begin{minipage}[h]{.7\textwidth}
    \begin{cppcode*}{}
    root [] namespace foo { }; struct S;
    root [] S *s; // Implicitly at ROOT startup
    root [] foo::bar /*store parsing state*/
    gSystem->Load("Foo");
    // More scaffolding.
    extern int __Cling_Autoloading_Map;
    namespace foo{struct __attribute__((annotate("$ClingAutoload$Foo.h"))) bar;}
    struct __attribute__((annotate("$ClingAutoload$Foo.h"))) S;
    // More initialization scaffolding.
    /*restore parsing state*/ *baz1;
    
    #include <Foo.h>/*restore parsing state*/;
    \end{cppcode*}
    \end{minipage}
    \caption{Information flow from {\it libFoo} dictionary.}
    \label{list:line2}
\end{listing}

Line 2 in Listing \ref{list:prompt} also does not require a definition. The second identifier lookup fails, but ROOT knows that {\it foo::bar} is in {\it libFoo} by the information from ROOTMAP files in Listing \ref{list:foo} Line 10. It dlopens {\it libFoo} which in turn, during its static initialization, inserts annotated forward declaration as shown in {\it G\_\_Foo.cxx}. This resolves {\it foo::bar} which avoids the parsing of {\it Foo.h} at relatively small overhead. The loading of the annotated forward declarations can happen at any time during parsing. This so-called "recursive parsing" is a code path that exists only in ROOT, and is not exercised by Clang itself. The behavior of Line 2 is equivalent to Listing \ref{list:line2} Lines 1 to 10.

Line 3 in Listing \ref{list:prompt} requires a definition and the implementation behaves exactly as in Line 2. When a definition is required, it reads the information in the annotation and also parses {\it Foo.h} as is shown in Line 12 in Listing \ref{list:line2}. The Line 3 in Listing \ref{list:prompt} behavior is equivalent to Line 1 to 12 in Listing \ref{list:line2}.

ROOTMAP files and RDICT files are important for ROOT's performance. However, they require various mechanisms to work together and can fail in corner cases. Also, due to the complicated way they are implemented on top of the Clang API, they have a high probability to become unusable if certain parts of Clang's internal behavior changes. C++ Modules can partly replace their performance benefits while also offering a more stable implementation.

\section{Implementation}
\label{implementation}

An implementation of the C++ Modules concept itself exists in the LLVM frontend Clang \cite{clang-Modules-doc} which works as an API for ROOT and is included in ROOT source tree. It is possible to compile ROOT with C++ Modules with other C++ compilers such as GCC, as it also compiles Clang in the source and ROOT calls its API. Clang supports the Modules TS and hosts Modules research and development work. Clang's implementation encourages incremental, bottom-up adoption of the C++ Modules \cite{Smith-cppcon}. The implementation is designed to work for C, C++, ObjectiveC, ObjectiveC++ and Swift \cite{Moduralize-doc}. Users can enable the Modules feature without modifications in header files. Clang allows users to specify Module interfaces in a dedicated file, called a Module Maps file. A Module Map file expresses the mapping between a Module file and a collection of header files. It can be mounted using the compiler’s virtual file system overlay mechanism to non-writable library installation paths. In practice, a non-invasive modularization can be done easily by introducing a Module Map file. In some cases the Module Map files can be automatically generated if the build system knows about the list of header files in every package.

Several steps were taken to adopt C++ Modules in ROOT. First, we enabled compilation of ROOT with C++ Modules, which improved ROOT's build time. This effort includes generating Module Map files and resolving cyclic header dependency inside ROOT. Next, we taught rootcling dictionary generator to generate PCM files attached with I/O information. It includes possibility for ROOT to preload all PCMS at the startup time in order to make declaration available without \#including the appropriate headers. Also, it was implemented the autoloading of libraries in ROOT which does not depend on old infrastructure (ROOTMAPS) and shows correctness benefits, described in section \ref{correctness} and is more efficient compared to ROOTMAPS.


For the C++ Modules adoption in CMS experiment \cite{cms}, we have been working closely with CMSSW team. As a result, ROOT was enabled with runtime C++ in CMS environment, and was implemented PCM generation for CMSSW libraries one by one. We could gradually migrate dictionary generation to PCM, as our current implementation falls back to ROOTMAP when a PCM is not generated, which enabled incremental migration from the old to the new infrastructure.

\subsection{Registration Mechanism and Automatic Discovery of C++ declarations}
\label{subsec:preloading}

\begin{listing}[h]
    \noindent
    \begin{minipage}[h]{.7\textwidth}
    \begin{cppcode*}{}
    root [] import ROOT.*;
    root [] import Foo.*;
    root [] foo::bar *baz1;
    \end{cppcode*}
    \end{minipage}
    \caption{Pseudocode shows the loading of all Modules at the ROOT startup time.}
    \label{list:implicit-include}
\end{listing}

A C++ Modules aware ROOT preloads all Modules at its startup time. Listing \ref{list:prompt} becomes equivalent to Listing \ref{list:implicit-include}. Listing \ref{list:implicit-include} shows the example of implicit \#include, where {\it foo::bar} can be used without even including {\it Foo.h}. With Modules, this feature is supported by importing all Modules at the startup time, as shown in Line 1 and 2. Currently, importing all Modules comes with a constant performance overhead which we explain in detail in section \ref{performance}.

However, there is another way to support the Global Modules Index feature, which generate the list of identifiers and PCMs at the initialization time. Upon the identifier lookup failure, the interpreter can refer to the list to decide which PCM to load. The possible drawback of this implementation is that the interpreter has no control over where the lookup failure can happen. It can happen inside a nested scope where importing a PCM may cause an incomprehensible error.

Regarding automatic discovery of C++ declarations, a naive implementation of this feature would require the inclusion of all reachable library descriptors at ROOT startup time, which is not feasible. ROOT inserts a set of optimizations from ROOTMAP files to fence itself from the costly full header inclusion. Unfortunately, several of them are home-grown and in a few cases inaccurate causing a notable technical debt.

In case of runtime C++ Modules, ROOT iterates through libraries found in Prebuilt Modules Paths and LD\_LIBRARY\_PATH until it finds the definition of the currently sought mangled name. It searches the library in Prebuilt Modules Path first as it is more likely that the symbols are in ROOT related library. When it fails, the implementation fall-backs to searching libraries from LD\_LIBRARY\_PATH for system library auto-loading. The overhead is remarkably low as it is only looking into a 64-byte hash in the library to determine whether this library likely has a definition or not, which is called Bloom Filter. Not only the symbols defined in regular symbol tables but also the dynamic symbols can be autoloaded as we are also checking .dynsym section where the dynamic symbols are defined. This feature is new and is not supported without C++ Modules in ROOT. The benefit of this implementation can be seen in Section \ref{correctness}.

\section{Results}
\label{results}

\subsection{Performance Results}
\label{performance}

\begin{figure}
\centering
    \begin{minipage}{.48\textwidth}
    \subfloat[] {\label{fig:perfBuildingROOT:a} \includegraphics[width=\textwidth]{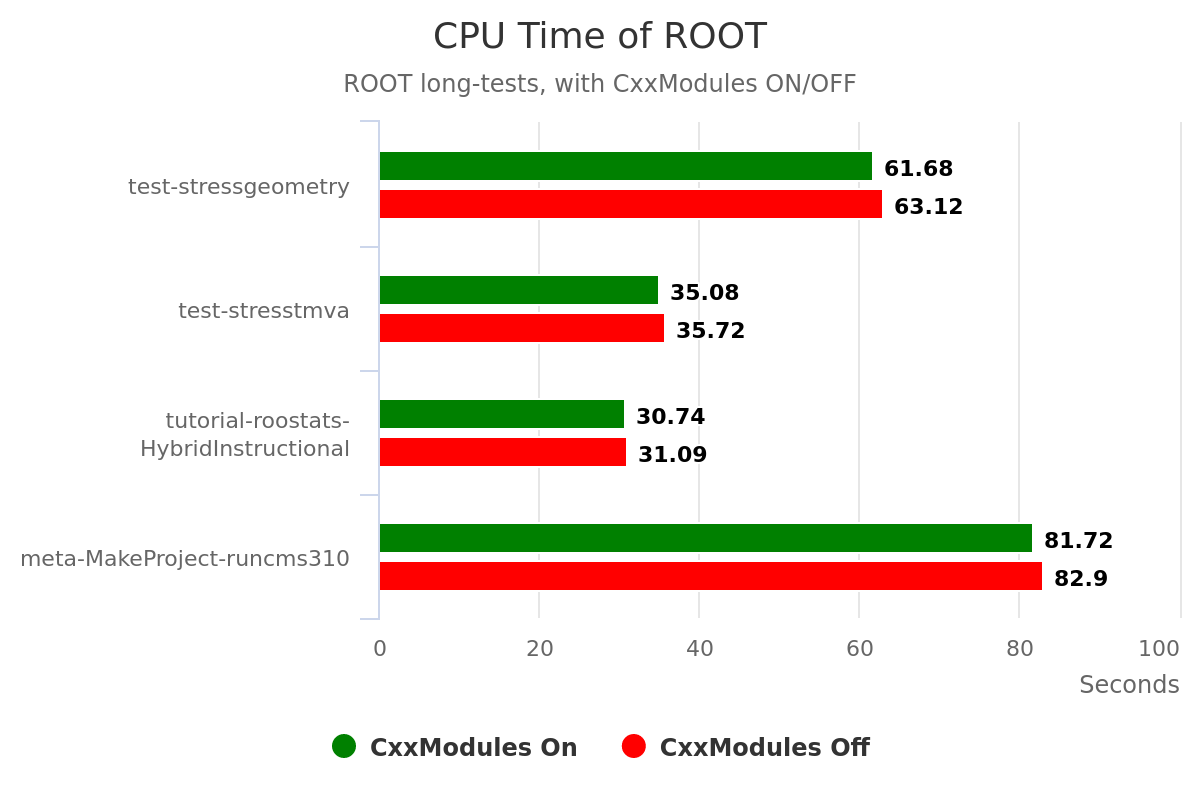}}
    \end{minipage}\hfill
    \begin{minipage}{.48\textwidth}
    \subfloat[] {\label{fig:perfBuildingROOT:a} \includegraphics[width=\textwidth]{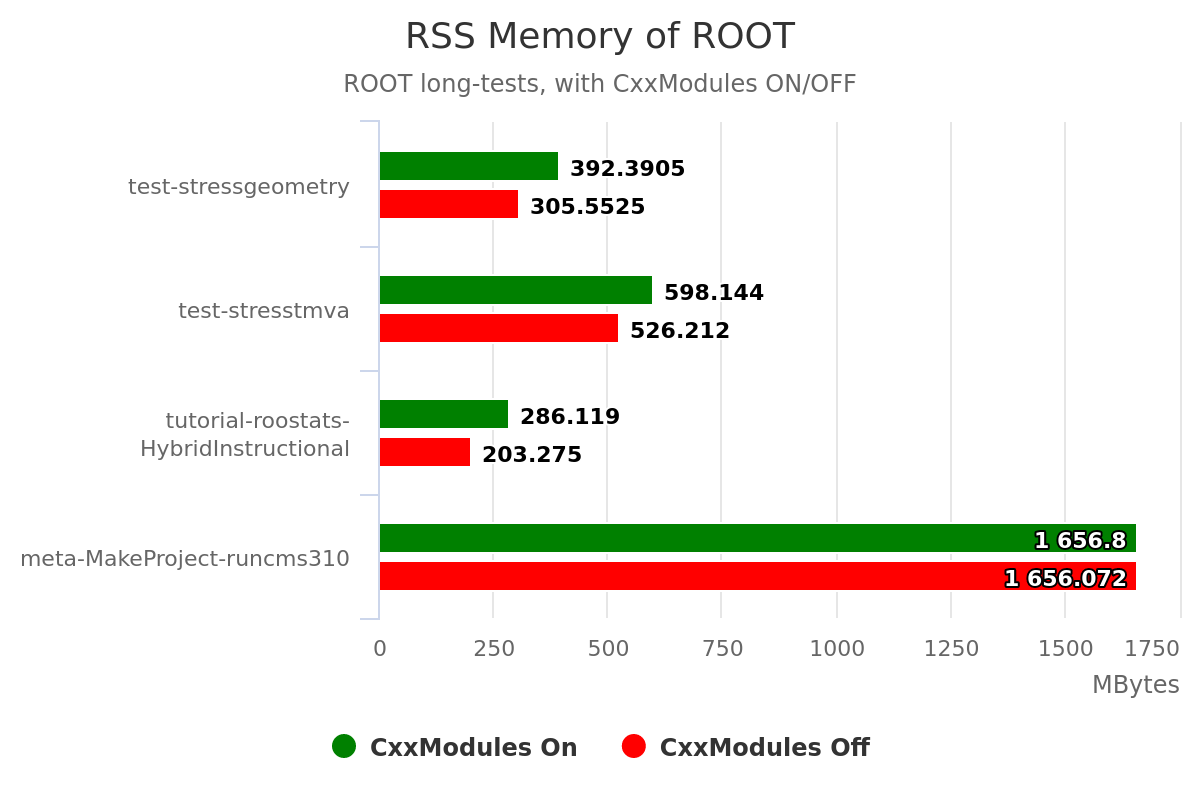}}
    \end{minipage}

    \begin{minipage}{.48\textwidth}
    \subfloat[] {\label{fig:perfBuildingROOT:a} \includegraphics[width=\textwidth]{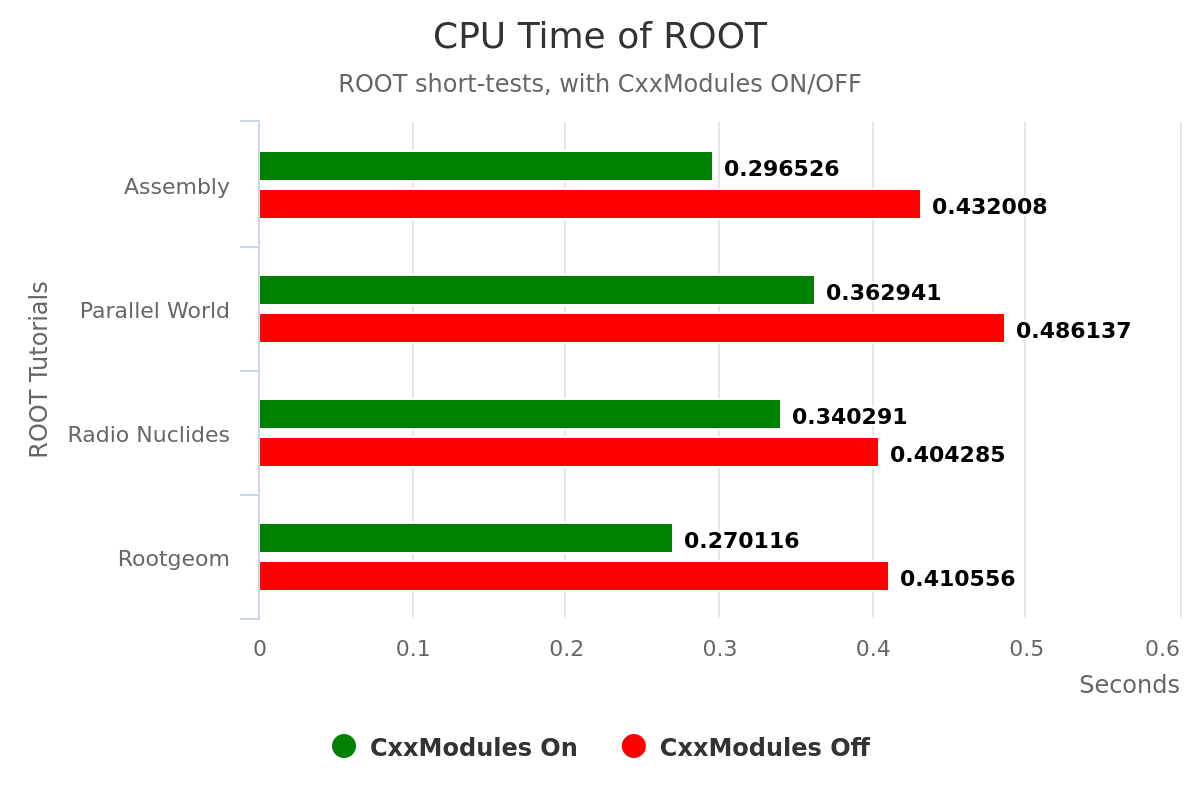}}
    \end{minipage}\hfill
    \begin{minipage}{.48\textwidth}
    \subfloat[] {\label{fig:perfBuildingROOT:a} \includegraphics[width=\textwidth]{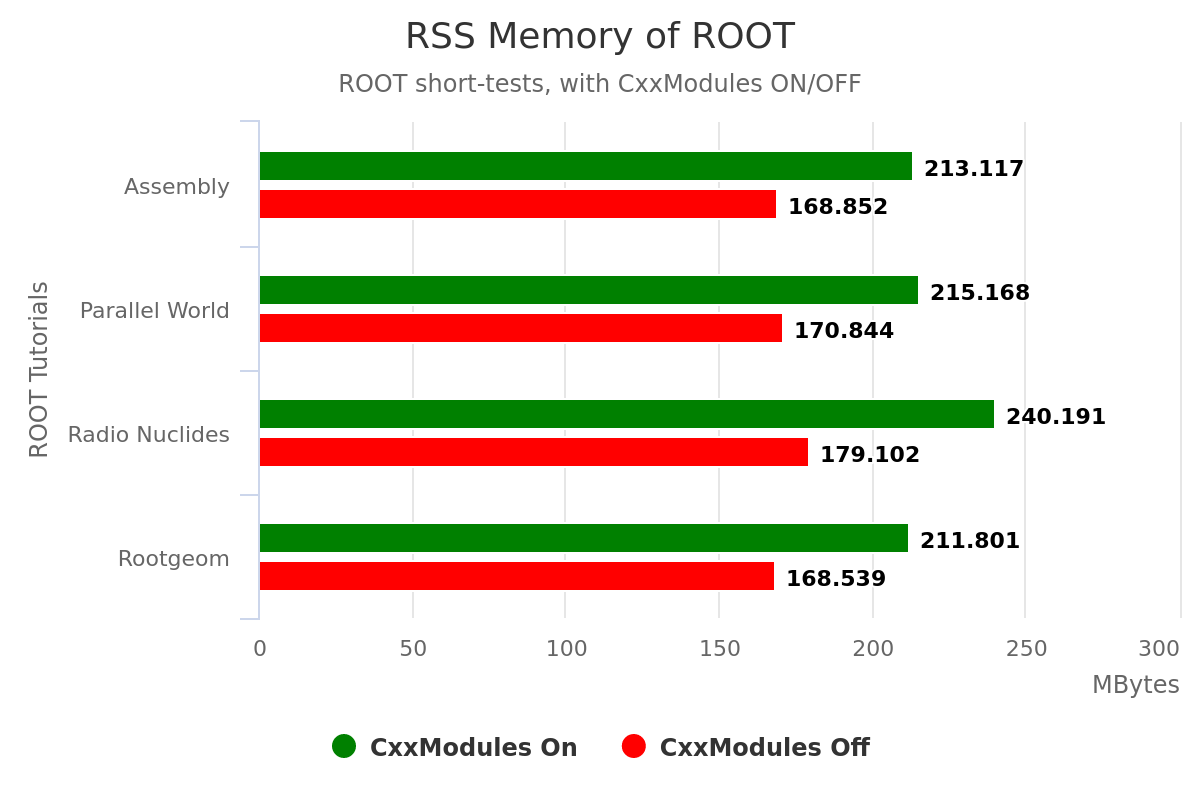}}
    \end{minipage}
\caption{Performance results: (a) and (c) are the measurement of CPU Time. (b) and (d) are the measurement of RSS. (a) and (b) are measuring long tests (over 30 seconds) in ROOT with and without runtime C++ Modules. (c) and (d) are measuring short tests which is not in PCH.}
\label{fig:performance1}
\end{figure}

\begin{figure}
\centering
    \begin{minipage}{.48\textwidth}
    \subfloat[] {\label{fig:perfBuildingROOT:a} \includegraphics[width=\textwidth]{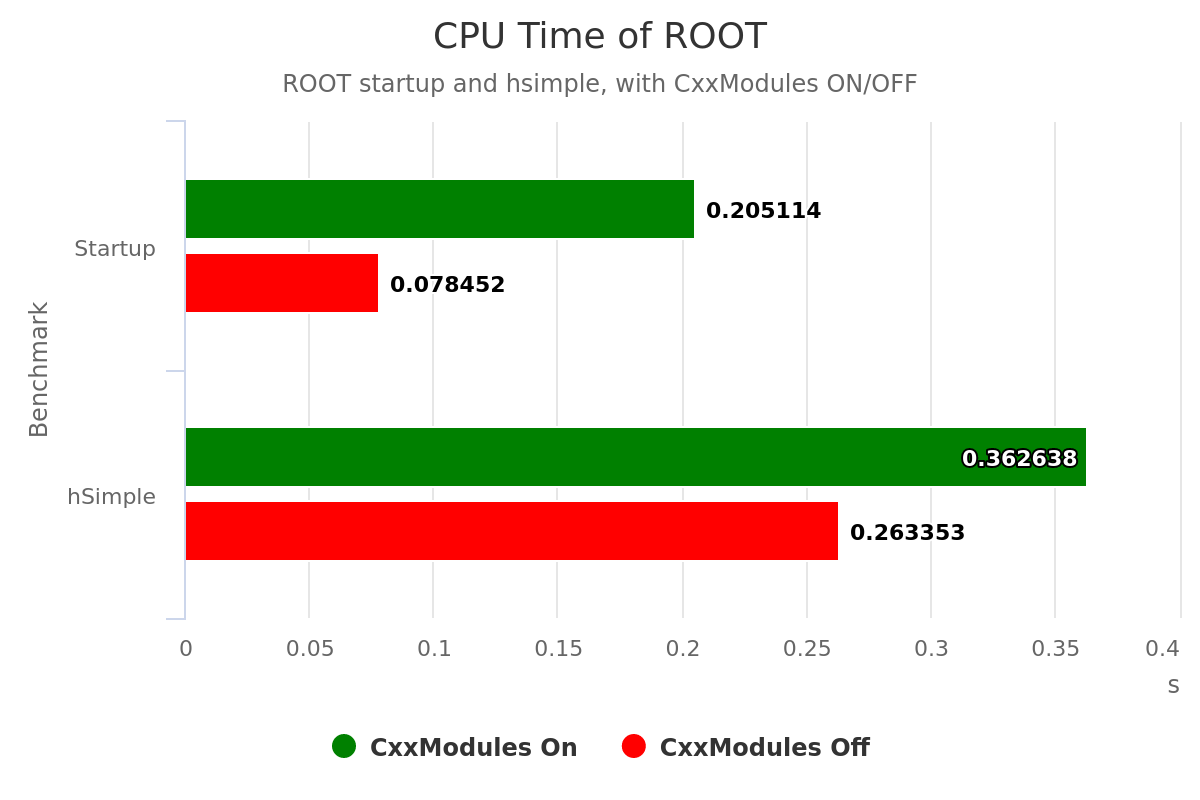}}
    \end{minipage}\hfill
    \begin{minipage}{.48\textwidth}
    \subfloat[] {\label{fig:perfBuildingROOT:a} \includegraphics[width=\textwidth]{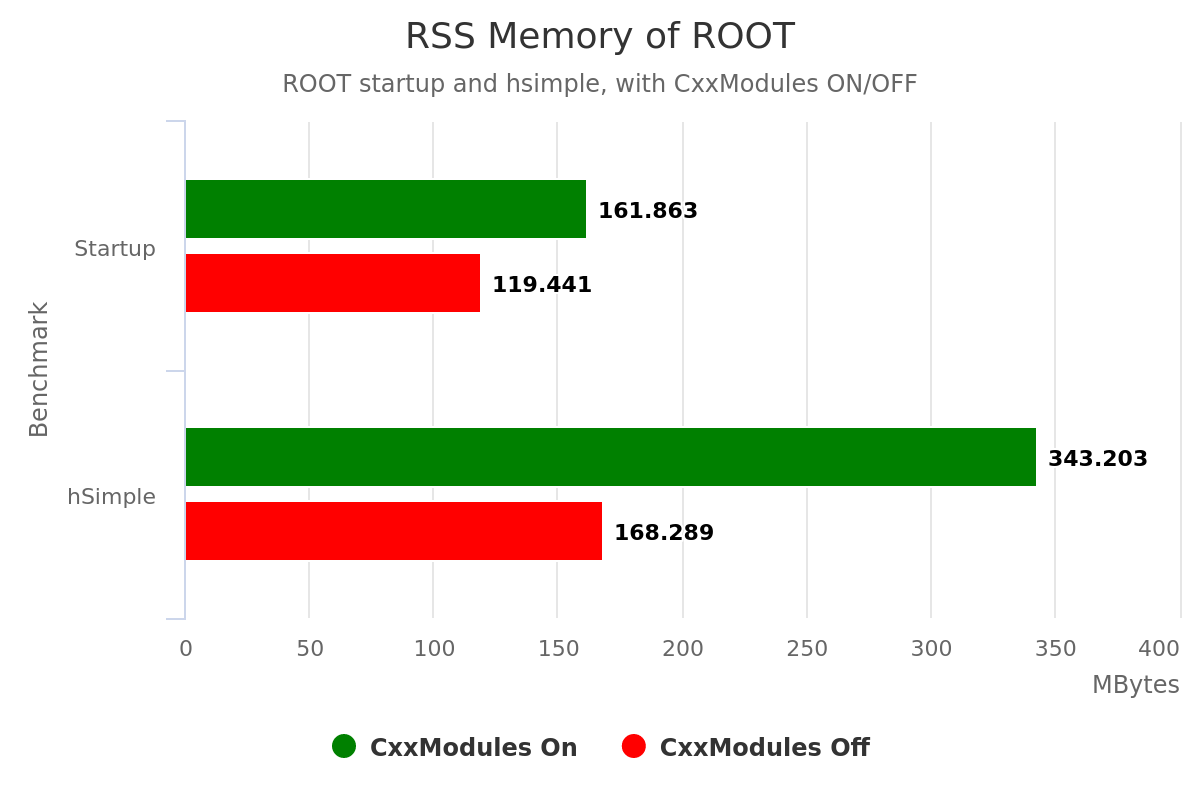}}
    \end{minipage}
\caption{Basic benchmarks: The start-up initialization of ROOT and hSimple tutorial.} 
\label{fig:performance2}
\end{figure}

ROOT Performance measurements are often done by running ROOT specific tutorials and tests. During tests, we measured the ROOT performance with {\it Archlinux 4.18.16 GNU/Linux, Intel(R) Core(TM) i7-8550U CPU} and {\it Ubuntu 18.04.1 LTS, i7-7500U NVIDIA GeForce 940MX}.
Figure \ref{fig:performance1} and Figure \ref{fig:performance2} shows the performance results for C++ Modules setup comparing to the PCH and Textual Headers setup, which are similar to the workloads of the experiment software stacks.

The ROOT CPU time is measured in (a) and (c) in Figure \ref{fig:performance1} and (a) in Figure \ref{fig:performance2}, whereas (b) and (d) in Figure \ref{fig:performance1} and (b) in Figure \ref{fig:performance2} are measuring the residential memory of ROOT.

ROOT long timing tests are the tests with duration of more then 30 seconds, measured in (a) and (b) in Figure \ref{fig:performance1}. (c) and (d) in Figure \ref{fig:performance2} are testing short tests which are not in PCH, which means that they are still using textual includes. Thus from those tests, we can get a rough assumption of the performance result we will get from modularizing experiments. Figure \ref{fig:performance2} is measuring the startup of ROOT and executing the hsimple tutorial, to show the actual startup time overhead we have from preloading PCMs.

The RSS memory regression, observed in Figure \ref{fig:performance1} (d) and in Figure \ref{fig:performance2} (b) are mostly due to procedure of importing all C++ Modules at the startup. The RSS overhead increases in the proportion to the number of the preloaded Modules. The startup memory overhead is between 40-60 MB depending on the concrete configuration. However, when the workload increases (Figure \ref{fig:performance1} (b)) we notice that the overall memory performance decreases in the number of cases.

The CPU time regression in Figure \ref{fig:performance2} (a) and in Figure \ref{fig:performance1} (c) are due to importing all PCMs at the startup time. However, in Figure \ref{fig:performance1} (a) the performance of C++ Modules is better than PCH by 1 or 2 seconds. It shows that C++ Modules can perform better when the workload of the users' code increases.

Performance of C++ Modules technology preview depends on multiple factors such as the ROOT configuration and a workflow organisation. We also implemented a continuous performance monitoring tool \cite{rootbench} where we compare the performance of the technology preview with respect to ROOT without C++ Modules.

\subsection{Correctness and Extra usability features}
\label{correctness}

\begin{listing}[h]
    \noindent
    \begin{minipage}[h]{.48\textwidth}
    \begin{cppcode*}{}
    root [] gMinuit // Cannot load variable
    IncrementalExecutor::executeFunction:
    symbol 'gMinuit' unresolved while
    linking [Cling interface function]!
    \end{cppcode*}
    \end{minipage}\hfill
    \begin{minipage}[h]{.48\textwidth}
    \begin{cppcode*}{}
    root [] gMinuit // Could load libMinuit
    (TMinuit *) nullptr
    \end{cppcode*}
    \end{minipage}
    \caption{Correctness results: The left-hand side is ROOT without runtime C++ Modules, which cannot autoload extern global variables such as gMinuit. The right-hand side is ROOT with runtime C++ Modules, with which gMinuit can be autoloaded.}
    \label{list:gMinuit}
\end{listing}

\begin{listing}[h]
    \noindent
    \begin{minipage}[h]{.48\textwidth}
    \begin{cppcode*}{}
    root [] #include <m17n-core.h> // System header
    root [] m17n_init_core()
    IncrementalExecutor::executeFunction:
    symbol 'm17n_init_core' unresolved while
    linking [Cling interface function]!
    \end{cppcode*}
    \end{minipage}\hfill
    \begin{minipage}[h]{.48\textwidth}
    \begin{cppcode*}{}
    root [] #include <m17n-core.h>
    root [] m17n_init_core()
    root [] // Autoload system library
    \end{cppcode*}
    \end{minipage}
    \caption{Autoloading of system libraries: The left-hand side is ROOT without runtime C++ Modules, which cannot autoload a system library. The right-hand side is ROOT with runtime C++ Modules, where ROOT can autoload the corresponding system library.}
    \label{list:system}
\end{listing}

As shown in Listing \ref{list:gMinuit}, gMinuit is an extern variable, that can't be autoloaded by ROOT without Modules since its forward declaration is not declared in ROOTMAP files. However, with Modules, we can automatically resolve symbols and cases like those will be correctly handled. Moreover, Listing \ref{list:system} shows that ROOT can also autoload system libraries with dynamic symbols. Module's autoloading implementation iterate through LD\_LIBRARY\_PATH which also includes system libraries. The implementation details are thoroughly discussed in Section \ref{implementation}.

\section{Limitations and Future work}
\label{limitationsandfuture}

Even though ROOT supports C++ Modules, there are still remained some issues to be solved before C++ Modules will become fully usable for developers and users. One of the limitations caused by fact that Clang does not explicitly support the relocation of implicitly-built PCM files. It is possible to patch Clang to work with implicitly-built PCM, but it is better to have an official support. The limitation comes from the fact that Modules files store paths of the configuration and source files in them. These paths will become invalid once the build directory has been moved.


One significant issue is that C++ Modules are currently not as efficient as ROOT's PCH when used in a minimal environment without experiment frameworks. In this situation the PCH is more efficient as it was optimized for the specific setup of ROOT. ROOT's C++ Modules however are kept back by two issues. First issue is an introduction of the additional overhead coming from management data structures that make PCMs more extensible than the PCH. Second issue is that C++ Modules in ROOT are still not well optimized comparing to the PCH. Especially the preloading of Modules on startup needs to be optimized as explained in section \ref{implementation}.

Our ultimate goal is to enable C++ modules as a default feature in ROOT. In order to achieve it, we will provide a support for adoption C++ Modules technology by experiments and continue optimizing the performance.

\section{Acknowledgments}

This work has been supported by an Intel Parallel Computing Center grant, by U.S. National Science Foundation grants PHY-1450377, OAC-1450377 and PHY-1624356, and by the U.S. Department of Energy, Office of Science.


\end{document}